\documentclass[runningheads]{llncs}
\usepackage[T1]{fontenc}
\usepackage{graphicx}
\usepackage{booktabs}
\usepackage{multirow}
\usepackage{xcolor}
\usepackage{wrapfig}

\usepackage{listings}
\usepackage{xcolor}

\lstdefinestyle{mcppython}{
language=Python,
basicstyle=\ttfamily\bfseries\fontsize{5.4}{6.0}\selectfont,
keywordstyle=\color{green!45!black}\bfseries,
stringstyle=\color{red!70!black},
commentstyle=\color{gray!70!black},
identifierstyle=\color{black},
columns=fullflexible,
keepspaces=true,
showstringspaces=false,
breaklines=true,
breakatwhitespace=false,
backgroundcolor=\color{black!5},
frame=none,
numbers=none,
aboveskip=2pt,
belowskip=2pt,
xleftmargin=3pt,
xrightmargin=3pt,
framexleftmargin=3pt,
framexrightmargin=3pt,
framextopmargin=3pt,
framexbottommargin=3pt,
tabsize=2
}

\usepackage{subcaption}
\usepackage{capt-of}
\usepackage{tabularx}
\usepackage{xspace}

\usepackage[
  colorlinks=true,
  linkcolor=red,
  citecolor=blue,
  urlcolor=blue
]{hyperref}

\newcommand{\yjyang}[1]{{\color{blue}{{[yjyang: #1]}}}}
\newcommand{\toolname}{\textsf{MCPPrivacyDetector}\xspace}

\begin{document}
\title{``What Happens Locally, Leaks Globally'': Detecting  Privacy Leakage Risks in MCP Servers}
\titlerunning{Detecting  Privacy Leakage Risks in MCP Servers}
\author{Biwei Yan\inst{1}\and 
Minghui Xu\inst{1} \and 
Yijun Yang\inst{1} \and
Boyang Ma\inst{1} \and
Xuelong Dai\inst{1} \and \\
Jinku Li\inst{2} \and
Yue Zhang\inst{1}}

\authorrunning{B. Yan et al.}
% \institute{Shandong University, Jinan 250100, China \\
% Xidian University, Xi'an 710071, China \\
% \email{bwyan@sdu.edu.cn, mhxu@sdu.edu.cn, ayyj.jun@gmail.com, boyangma@sdu.edu.cn, daixuelong@sdu.edu.cn, jkli@xidian.edu.cn, zyueinfosec@sdu.edu.cn}
% }

\institute{
Shandong University, Jinan 250100, China\\
\email{\{bwyan,mhxu,yijunyang,boyangma,daixuelong,zyueinfosec\}@sdu.edu.cn}
\and
Xidian University, Xi'an 710071, China\\
\email{jkli@xidian.edu.cn}
}
\maketitle              % typeset the header of the contribution
\begin{abstract}
The Model Context Protocol (MCP) has rapidly become the de facto standard for connecting large language models (LLMs) to external resources, but it also introduces a class of privacy risks that existing tools are ill-equipped to detect. Unlike conventional exfiltration bugs, leakage in MCP servers is largely \emph{protocol-induced}: credentials, API keys, and Personally Identifiable Information (PII) cross the local/LLM boundary simply by being returned, logged, or raised inside a tool handler, with no explicit outbound request in the source code.
We present \toolname, a context-aware cross-language static analysis framework that detects such leakage in multilingual MCP servers. \toolname lifts heterogeneous code implemented across different programming language (e.g., Python) into a unified program representation, applies context-aware semantic filtering to isolate genuinely sensitive values and protocol-specific implicit sinks (e.g., \texttt{@mcp.tool} handlers), and performs taint analysis to enumerate feasible flows. Applied to 10{,}655 real-world MCP servers, \toolname finds leakage rates above 10\%. Case studies confirm concrete exposures including leaked Bearer tokens, propagated API keys, and plaintext authentication credentials, arguing for systematic, protocol-aware safeguards in the emerging LLM agent toolchain.
\end{abstract}

\section{Introduction}
%\yjyang{The prior version is a bit wordy. I tightened the structure, removed the redundant explanations of MCP mechanics, and added a stronger framing of the architectural novelty of the risk}
Large language models (LLMs) are increasingly deployed as the reasoning core of autonomous agents that plan, invoke external tools, and operate on real-world resources~\cite{wu2024autogen}. To support this shift, the community has rapidly converged on the Model Context Protocol (MCP) as a de facto standard for connecting agents to heterogeneous backends such as local file systems, private databases, and third-party APIs~\cite{anthropic2024mcp,mcpdocs2025intro,hou2025model}. By standardizing context exchange and tool invocation, MCP has enabled a vibrant ecosystem of \emph{MCP servers}, in which each server exposes a set of domain-specific tools that the LLM can call during task execution.

The MCP server sits at the boundary between local computation and remote inference. It reads from the local environment, runs tool logic, and serializes results into a protocol-conformant response that the MCP client forwards to the LLM as part of its context window~\cite{hasan2026model,radosevich2025mcp}. What appears to be a simple request--response pattern is in fact a chain of implicit data flows: function return values are automatically captured by protocol decorators (e.g., \texttt{@mcp.tool}), exceptions surface as structured error payloads, and logging side-effects leak into both local storage and the outbound response. In this setting, the boundary between \emph{debugging artifact} and \emph{model context} becomes dangerously thin, raising risks around information leakage.

This paper studies a class of privacy risks that arise directly from that architectural mismatch. Unlike conventional data-exfiltration bugs~\cite{aghili2025protecting,gu2022logging,han2023credential,li2014know}, which require an explicit outbound request, privacy leakage in MCP servers is largely \emph{protocol-induced}: sensitive values enter the leakage path simply by being returned, logged, or raised inside a tool handler. We observe two concrete manifestations. First, verbose debugging practices cause credentials, API keys, and personally identifiable information (PII) to persist in plaintext within local logs and consoles, where they remain accessible to anyone with host-level access. Second, and more insidiously, such values are silently folded into serialized tool outputs and error messages; once the client forwards them, the remote LLM treats them as legitimate context and may echo them to downstream services, cache them, or surface them in later responses. The net effect is that the sensitive local state is exfiltrated without any explicit network call appearing in the server's source code.

Detecting these leaks is non-trivial for three reasons. \emph{(i) Implicit sinks.} MCP's decorator- and callback-driven design hides the actual data-egress boundary behind framework-level machinery, which pattern-matching scanners and conventional taint tools do not model. \emph{(ii) Context-sensitive sensitivity.} Whether an identifier such as \texttt{user\_id} or \texttt{token} denotes secret material depends on surrounding semantics; benign uses (e.g., \texttt{tokens\_used} metrics) routinely trigger false positives in na\"ive rule-based detectors. \emph{(iii) multi-language fragmentation.} Production MCP servers are written in Python, Go, Java, JavaScript, and TypeScript, each with its own AST shape, idioms, and framework bindings, making cross-ecosystem analysis prohibitively expensive with per-language tooling.

We present \toolname, a static analysis framework purpose-built for privacy-leakage detection in the multi-language MCP ecosystem. \toolname makes three design choices that jointly address the challenges above. First, it lifts heterogeneous source code from Python, Go, Java, JavaScript and TypeScript into a \emph{unified relational program representation}, normalizing cross-language AST differences into a common query substrate. Second, it applies \emph{context-aware semantic filtering} that combines lexical cues (configuration files, hardcoded credentials, PII-shaped identifiers) with AI-specific contextual signals (e.g., token-accounting fields and length-check idioms) to separate genuinely sensitive values from superficially similar but benign ones, while simultaneously locating protocol-specific sinks such as \texttt{@mcp.tool} handlers, logging calls, and exception surfaces. Third, on top of this model, \toolname performs inter-procedural \emph{taint analysis} to enumerate feasible flows from sensitive sources to MCP-visible sinks, yielding an automated verdict on whether a given server is exposed to privacy leakage.

We evaluate \toolname on 10{,}655 real-world MCP servers collected from GitHub and the major public MCP registries. The results show that privacy leakage is already endemic: leakage rates exceed 10\% in every language we study, peaking at 19.1\% in Java and 15.4\% in Python, and 83.0\% of all detected cases concentrate in three registries, namely Pulse MCP, Smithery Registry, and Mcpmarket. Case studies confirm that these findings translate directly into exploitable risk. Bearer tokens surfaced in server logs permit direct credential replay; API keys embedded in outbound HTTP headers carry private data beyond the local execution boundary; and authentication secrets printed to standard output propagate into downstream log-collection pipelines. Together, these results indicate that MCP privacy leakage is not an isolated defect but a systemic weakness, concentrated in a small number of ecosystems and a handful of recurring patterns.

\smallskip
\noindent\textbf{Contributions.} This paper makes the following contributions:
\begin{itemize}
    \item \textbf{A new class of privacy risk.} We identify and characterize a privacy-leakage class specific to MCP, rooted in the semantic mismatch between local tool execution and remote context propagation, whereby return values, exceptions, and debug output implicitly cross the local/LLM boundary.

    \item \textbf{A multi-language detection framework.} We present \toolname, a static analysis framework that combines a unified cross-language program representation, context-aware semantic filtering, and protocol-guided taint analysis to detect such leaks at scale across Python, Go, Java, JavaScript, and TypeScript servers.

    % \item \textbf{A large-scale ecosystem study.} We conduct, to our best knowledge, the first systematic security analysis of MCP servers across five languages and 10{,}655 repositories, quantifying the prevalence of privacy leakage and exposing concrete gaps in the privacy safeguards of emerging LLM agent toolchains.

    \item \textbf{A large-scale ecosystem study.} We conduct a large-scale security analysis of MCP servers across five languages and 10{,}655 repositories, quantifying the prevalence of privacy leakage and exposing concrete gaps in the privacy safeguards of emerging LLM agent toolchains.
\end{itemize}
\section{Background and Motivation}

\subsection{Execution Workflow in MCP}\label{workflow}

\begin{figure}[t]
  \centering
  \includegraphics[width=0.8\linewidth]{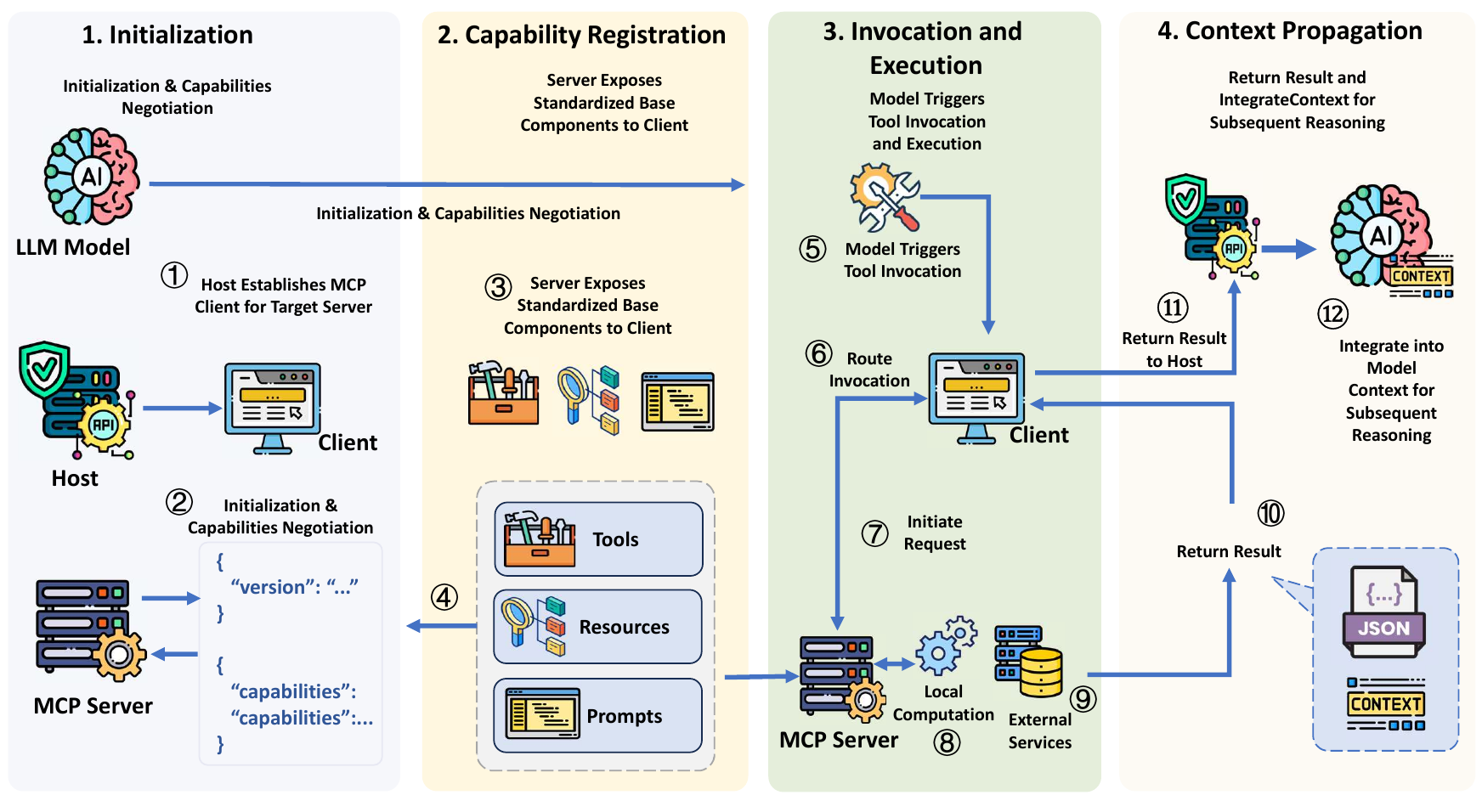}
  \caption{Execution in MCP}
  % \caption{Execution workflow in MCP. A tool invocation traverses four stages: the host initializes a client session with the server~(1), the server registers its tools, resources, and prompts via protocol decorators~(2), the LLM-triggered call is dispatched to a local handler whose return value is serialized~(3), and the result is propagated back into the model's context window~(4). Stage~(4) is the implicit egress path exploited by the privacy leakage studied in this paper.}
  \label{img:mcp_flow}
\end{figure}

MCP is an open protocol that standardizes how LLM-based applications interact with external capabilities such as tools, data resources, and prompt templates~\cite{anthropic2024mcp,mcpdocs2025intro,hou2025model}. It adopts a \texttt{host--client--server} architecture: the host embeds the LLM, each client maintains a stateful session with one MCP server, and each server exposes a set of local capabilities to the agent. Figure~\ref{img:mcp_flow} illustrates the four stages of a typical invocation.
\noindent\textbf{(1) Initialization.} The host instantiates a client, negotiates protocol version and transport with the target MCP server, and establishes a session over which subsequent messages are exchanged.
\noindent\textbf{(2) Capability registration.} The server advertises the components it exposes tools, resources, and prompts, through a standardized schema. In practice, these are typically bound via framework-level decorators (e.g., \texttt{@mcp.tool}) that transparently wire local functions to protocol endpoints. 
\noindent\textbf{(3) Invocation and execution.} When the LLM decides to call an external capability, the host forwards the call through its client to the MCP server; the server dispatches it to the corresponding handler, executes the handler against the local environment, and serializes the result into a protocol-conformant response.
\noindent\textbf{(4) Context propagation.} The client returns the serialized result to the host, which incorporates it into the LLM's context window as a tool-use observation, closing the loop so the model can proceed with the task.

This design gives MCP its flexibility: any backend that speaks the protocol becomes addressable by any MCP-aware agent. The same design, however, makes the boundary between local execution and the LLM's context implicit---stage~(4) silently transports whatever stage~(3) produced, including values that were never intended to leave the host. 

%As discussed in \S\ref{workflow}, MCP servers act as middleware between LLM-based agents and external resources, including local configuration files, user-provided inputs, third-party APIs, and tool execution results. Since these resources may involve credentials, personal information, or other privacy data during access, processing, and return, the MCP server may cause privacy data leakage when processing these information. To determine whether such privacy-relevant information exists in real-world MCP implementations, we analyzed the privacy data sources identified in our dataset and summarized their category distribution in Figure \ref{img:MCP_PII}. The results indicate that data containing privacy-sensitive information is prevalent within MCP servers. Specifically, credentials and API keys constitute the dominant category, followed by contact information, while identity-related, financial, and address-related information also appear in non-negligible quantities. These findings indicate that MCP servers do not merely process operational metadata. Rather, they frequently embed and manipulate diverse forms of privacy-sensitive data as part of normal tool execution, request handling, and configuration management. In summary, existing MCP servers unintentionally expose a high risk of privacy leakage.

\subsection{Motivation and Problem statement}
\subsubsection{Motivation.} Although prior studies have already suggested that privacy risks may exist in LLM agents~\cite{yi2025benchmarking,zhan2024injecagent}, privacy leakage in MCP servers remains underexplored, despite their central role as a middleware~\cite{zhang2024privacy,wang2025unveiling,debenedetti2024agentdojo}. To better understand this problem, we conducted an empirical study on real-world MCP servers. Specifically, we collected 10,655 MCP servers and analyzed the privacy-related information. Our analysis showed that 6,657 servers contained privacy-related information, accounting for 62.5\% of the total. Furthermore, 1,317 servers exhibited privacy leakage risks, representing 12.4\% of all servers and 19.8\% of those containing privacy-related information. We further identified privacy-relevant entities and categorized them semantically. Credentials constituted the dominant category (56.22\%), followed by contact information (29.78\%), while identity-related, financial, and address-related information also appeared in non-negligible quantities. These findings indicate that MCP servers frequently embed and manipulate privacy-sensitive data during tool execution, request handling, and configuration management, thereby creating substantial opportunities for unintended exposure. Therefore, the core problem is how to accurately identify privacy leakage risks in MCP servers.

Privacy leakage in MCP servers manifests through two complementary mechanisms, both rooted in the workflow of Section~\ref{workflow}.

\vspace{5pt}
\noindent\textbf{(I) Local plaintext leakage.} The execution chain of an LLM agent is opaque by construction: a single user query can trigger multiple interleaved tool calls whose intermediate state is invisible from the model side. To regain observability, developers instrument MCP servers with verbose runtime logging, stack traces, and temporary caches. The same handlers, however, routinely process sensitive material such as OAuth tokens, API keys, and PII, embedded in request payloads and exception contexts during stage~(3) of the invocation flow. When these values are written in plaintext to log files, standard output, or on-disk caches, they persist on the host beyond the lifetime of the invocation and become accessible to any process, collaborator, or log-aggregation pipeline with host-level access.

\begin{wrapfigure}{r}{0.45\linewidth}
  \centering
  \includegraphics[width=\linewidth]{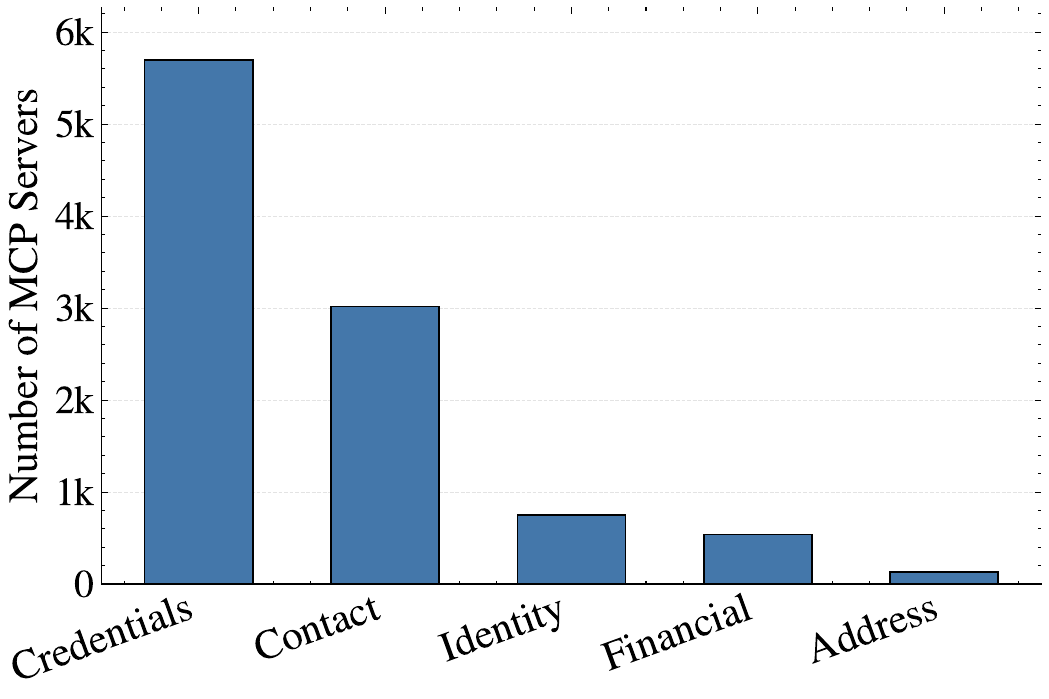}
  \caption{Distribution of privacy information in MCP servers by category}
  \label{img:MCP_PII}
\end{wrapfigure}

\vspace{5pt}
\noindent\textbf{(II) Implicit cross-boundary propagation.} A second, more insidious mechanism is specific to MCP's protocol design. Because stage~(4) serializes whatever a tool handler returns and forwards it into the LLM's context window, the absence of handler-level sanitization means any sensitive value surfaced during execution, whether as a legitimate field of the return object, an unhandled exception message, or a debug detail the developer forgot to strip, is transmitted to the remote model verbatim. From the protocol's perspective, these values are indistinguishable from intended context: they are encapsulated as legitimate tool-use observations, cross the local/LLM boundary implicitly, and may subsequently be logged, cached, or echoed to downstream services by the model host. The result is data exfiltration without any explicit outbound request in the server's source code, which is precisely what makes the risk easy to overlook during development and code review.

The two mechanisms compound. A value leaked into a log is frequently also surfaced in the corresponding return object, and a value returned to the LLM is often logged for debugging, so a single handling mistake can simultaneously expose the same secret across local and remote surfaces.

\vspace{5pt}
\noindent\textbf{Problem statement.}  We formulate privacy leakage detection in MCP servers as a source-to-sink analysis problem. In an MCP server, privacy-sensitive information may obtained from various program entities, such as hardcoded credentials, sensitive configuration files, environment-dependent values, or variables and fields that carry privacy-relevant semantics. A privacy leak occurs when such data can propagate through the server logic and reach externally observable endpoints, including protocol-visible outputs, logging operations, or outbound communication interfaces. Accordingly, our goal is to determine whether an MCP server contains feasible privacy leakage paths from sensitive sources to leakage-relevant sinks.

\section{Design of \toolname }

\subsection{Challenge and and solution}
\vspace{5pt}
\noindent\textbf{C1. Multi-language heterogeneity in MCP server.}
In the MCP ecosystem, MCP is an open protocol that standardizes interactions between LLM applications and external data sources and tools, while remaining independent of any specific implementation language. Accordingly, official MCP documentation and SDKs support server development in multiple languages, including Python, JavaScript, Java,   Go and TypeScript~\cite{mcp_spec_2025_03_26}. As a result, the same privacy-relevant behavior may appear in very different program forms across implementations. This heterogeneity makes it difficult to analyze MCP servers directly over raw source code and limits the applicability of existing single-language techniques. Therefore, an effective solution needs to provide a unified cross-language abstraction that can standardize heterogeneous program structures while preserving privacy-related features for subsequent analysis.

\vspace{5pt}
\noindent\textbf{C2. Ambiguous leakage metrics and benign contexts.}
Privacy leakage detection in MCP servers generally uses rule-based methods for identifying sensitive features, such as hardcoded keys, sensitive identifiers, file accesses, logging statements, HTTP request parameters, and externally visible return values. However, prior studies show that pattern-based detection is highly prone to false positives~\cite{basak2023comparative},  since many benign artifacts may look similar to sensitive data. For example, variables containing keywords such as \textit{token}, \textit{key}, or \textit{address} may simply denote telemetry counters, file paths, protocol fields, or resource locators, while expressions such as \texttt{tokens\_used}, \texttt{max\_tokens}, \texttt{len(secret)}, \texttt{hex()}, or \texttt{repr()} do not necessarily imply real leakage of sensitive information. Therefore, privacy leakage detection in MCP servers cannot rely solely on isolated lexical or structural features; it requires context-aware semantic filtering to distinguish genuine privacy risks from benign program behavior.

\vspace{5pt}
\noindent\textbf{C3. Detecting privacy leakage under implicit data propagation.}
Even when sensitive features are present in a program, their existence alone does not imply a privacy leakage~\cite{gordon2015information}. The real risk lies in whether sensitive data is propagated through assignments, parameter passing, or attribute accesses, and eventually reaches observable leakage endpoints, such as logging outputs, outbound HTTP request parameters, or tool return values. For MCP servers, this issue is particularly pronounced because sensitive information is rarely exposed directly at its point of definition. Instead, it is often propagated through local assignments, helper-function encapsulation, object field accesses, and intermediate computations. Eventually, such sensitive data may be embedded into protocol context or transmitted to potential leakage sinks (e.g., external communication interfaces), thereby resulting in privacy leakage. Therefore, we must not only identify sensitive features but also further verify whether this data can propagate along actual program paths to potential leakage points. Otherwise, it is difficult to accurately determine whether the MCP server truly poses a risk of privacy leakage. Thus, we need an inter-procedural data-flow analysis method to determine whether sensitive data is actually leaked.

\begin{figure}[t]
  \centering
  \includegraphics[width=0.8\linewidth]{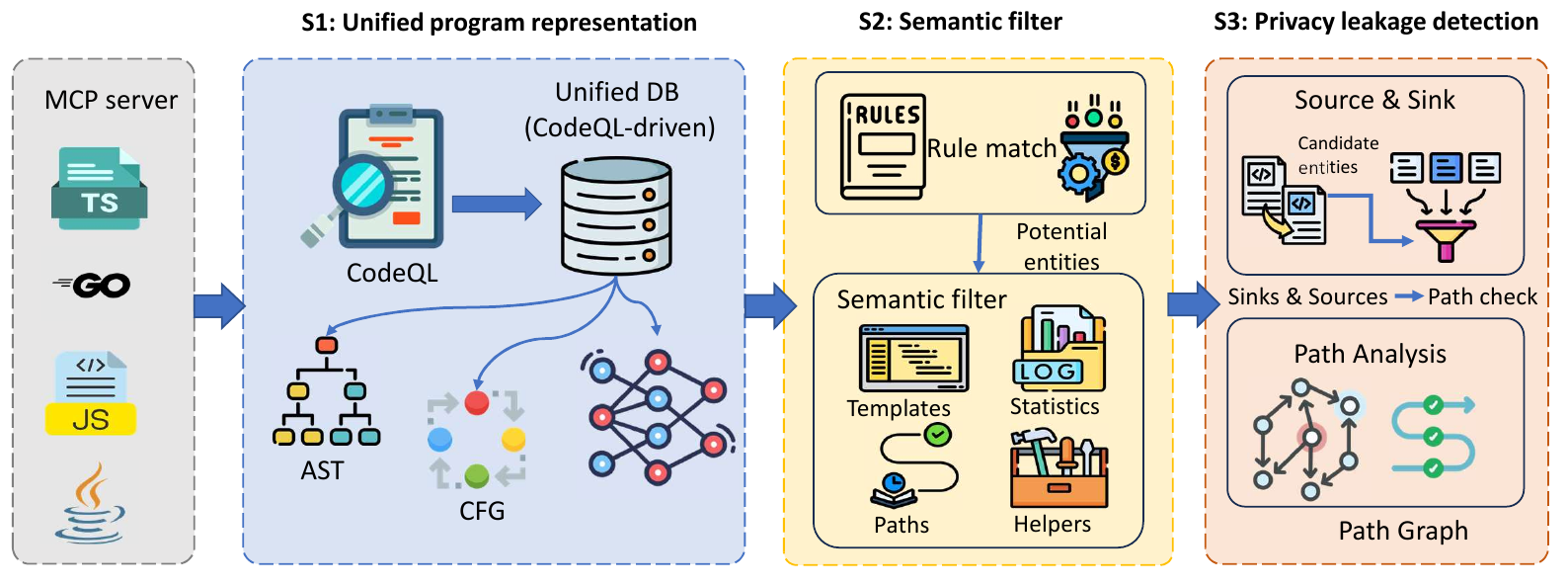}
  \caption{Overview of \toolname }
  \label{img:mcp_overview}
\end{figure}

To address these challenges, we propose \toolname , a context-aware static analysis framework designed for detecting privacy leakage in MCP servers as shown in Figure \ref{img:mcp_overview}.  This framework comprises three core components: a unified cross-language program representation, context-aware rule matching and filtering, and taint-analysis-based leakage detection.  Specifically, the framework first establishes a unified cross-language analysis foundation to eliminate differences in syntax and program structure across different implementation languages, providing a consistent program abstraction and thus improving the applicability of the method in a multi-language MCP ecosystem. Building on this, unlike previous methods that relied solely on isolated pattern matching or lexical scanning leading to high false positives, the framework introduces a context-aware semantic filtering mechanism. This extracts candidate entities with privacy-related features to exclude a large number of benign matches that are similar to sensitive patterns, thereby reducing false positives. Finally, the framework uses taint analysis to detect the actual propagation relationships between candidate entities, thereby determining whether the MCP server truly poses a risk of privacy leakage.

\vspace{5pt}
\noindent\textbf{S1. CodeQL-based unified multi-language program representation.} To solve the challenge regarding the heterogeneity of programming languages in MCP servers (Challenge C1), the framework first uses CodeQL to establish a cross-language foundation for static analysis. It translates MCP servers written in various languages into a unified and queryable program representation. Subsequent analyses no longer rely on language-specific syntax but are instead operate on unified program entities and their interrelationships. In this manner, the framework transforms the problem of detecting privacy leakages in multi-language MCP servers into a CodeQL-based program analysis problem, thereby providing a common foundation for rule matching and data-flow analysis.

\vspace{5pt}
\noindent\textbf{S2. Rule matching and context-aware semantic filtering.} To address the issue of false positives identified in Challenge C2, \toolname  employs a strategy that combines heuristic rule matching with context-aware semantic filtering. First, it extracts candidate entities that may carry privacy-related semantics. Then, it examines the program context surrounding these entities to filter out entities irrelevant to privacy information, such as template constants, statistical indicators, path descriptions, or matches corresponding to internal auxiliary logic. This process effectively eliminates noise from the initial set of candidate entities, thereby enhancing the accuracy of the detection results.

\vspace{5pt}
\noindent\textbf{S3. Taint analysis-based privacy leakage detection.} As discussed in Challenge C3, the presence of sensitive entities only indicates that a program may contain privacy-related information; it does not directly imply that a privacy leakage has actually occurred. A risk of privacy leakage exists only when such data can propagate along actual program execution paths to reach an observable output boundary. To address this issue, \toolname  further introduces a privacy leakage detection scheme based on taint analysis. The initial candidate entities generated in the previous stage are formally designated as sources and sinks. Then, a check is performed to determine whether an actually reachable data flow path exists between them. If such a path is indeed present, the analyzed MCP server is deemed to be at risk of privacy leakage.
\subsection{Detailed Design}\label{design}
\subsubsection{Cross-language program representation and entity modeling.}

\toolname uses CodeQL to convert source code from different languages into a queryable program representation. 

\vspace{5pt}
\noindent\textbf{Step 1: Cross-language code parsing and database construction.}
In the first step, the framework creates a CodeQL database for each MCP server. For each project, CodeQL parses the source code and converts it into a language-aware but queryable program representation. This database captures structural and semantic information such as \texttt{identifiers}, \texttt{attribute accesses}, \texttt{function calls}, \texttt{assignments}, \texttt{return values}, and \texttt{data-flow relations}. Although different languages (e.g., Python, JavaScript, Go) differ substantially in syntax and execution behavior, CodeQL lifts them into a common queryable form. 

\vspace{5pt}
\noindent\textbf{Step 2: Program entity extraction and organization.}
Based on the constructed database, the framework then organizes the key program entities required for privacy leakage detection. Specifically, the rules (see Table \ref{tab:sensitive_rules}) are built on \texttt{identifier nodes}, \texttt{attribute nodes}, \texttt{call nodes}, \texttt{assignment}\texttt{-related} \texttt{ statements}, and \texttt{return expressions}. These entities provide the basis for modeling sensitive sources, potential sinks, and intermediate propagation paths. In this way, the detection task is formulated as a relation analysis problem over structured program entities extracted from the CodeQL database.

\subsubsection{Rule matching and context-aware filtering.}

\begin{table*}[t]
\centering
\caption{Six rules for privacy leakage detection in \toolname}
{
\begingroup
\setlength{\tabcolsep}{2.0pt}
\renewcommand{\arraystretch}{0.92}
\fontsize{8}{7.2}\selectfont
\resizebox{0.98\textwidth}{!}{%
\begin{tabular}{p{0.9cm}p{3.2cm}p{4.2cm}p{4.8cm}p{3.8cm}}
\toprule
\textbf{Rule ID} & \textbf{Rule Category} & \textbf{Trigger Condition} & \textbf{Semantic Constraint} & \textbf{Detection Role} \\
\midrule
L1 & Hardcoded Secret Literal &
A variable or attribute is assigned a string literal &
The identifier carries explicit secret semantics, such as \texttt{secret}, \texttt{token}, \texttt{api\_key}, or \texttt{access\_key} &
Identifies hardcoded secret or credential literals and models them as high-confidence taint sources. \\
\midrule
L2 & PII Cleartext Logging &
A print or logging operation directly outputs a variable or expression &
The printed or logged content carries general PII semantics, such as \texttt{email}, \texttt{phone}, or \texttt{address} &
Identifies plaintext exposure of ordinary PII through local output operations and models them as leakage sinks. \\
\midrule
L3 & Sensitive Cleartext Logging &
A print or logging operation directly outputs a variable or expression &
The printed or logged content carries stronger secret semantics, such as \texttt{token}, \texttt{secret}, \texttt{password}, or \texttt{api\_key} &
Identifies plaintext exposure of sensitive credentials or secrets through local output operations and models them as leakage sinks. \\
\midrule
L4 & Sensitive File Read &
A local file is opened and its contents are subsequently read &
The file path indicates sensitive semantics, such as \texttt{.env}, \texttt{mcp.json}, \texttt{id\_rsa}, certificate files, or common credential/configuration files &
Identifies data loaded from sensitive local files and models them as taint sources. \\
\midrule
L5 & Sensitive File Flow to External Sink &
Data originating from a sensitive file flows to an external sink &
The destination corresponds to an externally observable sink, such as an outbound HTTP request, a tool return value, or another externally exposed output location &
Identifies cross-boundary leakage paths in which sensitive file content is propagated to external sinks. \\
\midrule
L6 & Sensitive Data Flow to External Sink &
A sensitive variable or attribute flows to an external sink &
The propagated data carries sensitive semantics and reaches an externally observable sink, such as an outbound HTTP request, a tool return value, or another externally exposed output location &
Identifies cross-boundary leakage paths in which sensitive in-memory data is propagated to external sinks. \\
\bottomrule
\end{tabular}%
}
\endgroup
}
\label{tab:sensitive_rules}
\end{table*}

In this stage, the framework first extracts source and sink candidates using rule-based static matching, and then refines them through context-aware filtering over the CodeQL representation, providing reliable inputs for subsequent taint analysis.

\begin{wrapfigure}{r}{0.45\linewidth}
  \centering
  \includegraphics[width=\linewidth]{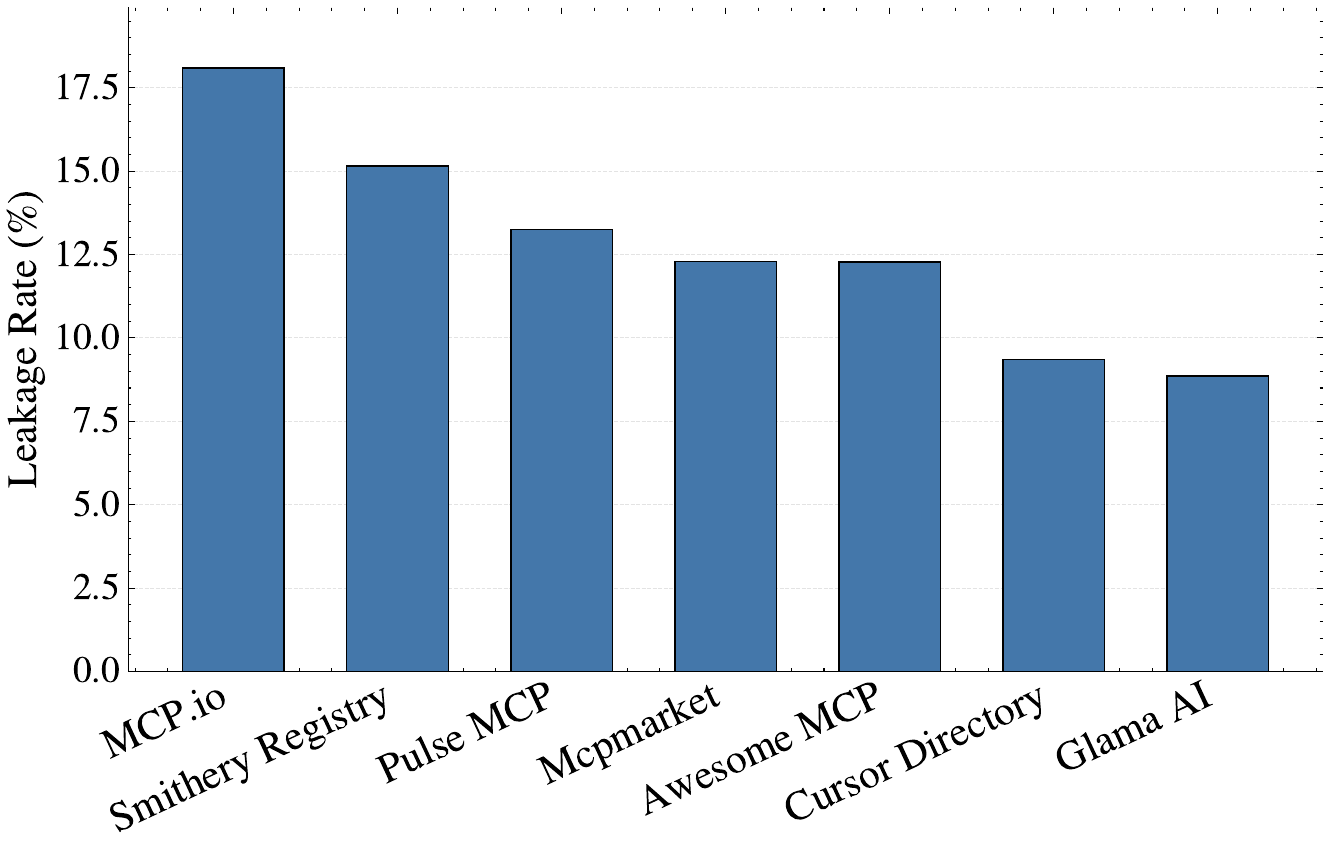}
  \caption{Privacy leakage rate across MCP markets}
  \label{img:MCP_PII_leakage_rate_by_market}
\end{wrapfigure}

\vspace{5pt}
\noindent\textbf{Step 1: Rule-based candidate extraction.}
The framework defines six privacy-leakage detection rules, as summarized in Table~\ref{tab:sensitive_rules}. These rules capture three complementary aspects of leakage analysis: \emph{source identification}, \emph{local plaintext exposure}, and \emph{cross-boundary propagation}. Specifically, \emph{hardcoded secret literal} and \emph{sensitive file read} identify high-confidence sensitive sources, while \emph{PII cleartext logging} and \emph{sensitive print or log} capture local plaintext exposure through print or logging operations. In contrast, \emph{sensitive file flow to HTTP or tool} and \emph{sensitive flow to HTTP or tool} characterize end-to-end propagation paths in which sensitive data reaches externally observable sinks, such as outbound HTTP requests or tool-returned outputs. Each rule matches candidate program entities or flows through recognizable structural patterns, such as literal assignments, file-read operations, output-related calls, and source-to-sink propagation. The result of this step is an initial set of candidate sensitive sources, local exposure points, and external leakage paths for later analysis.

\vspace{5pt}
\noindent\textbf{Step 2: Context-aware candidate refinement.}
Since the candidates extracted in Step 1 are intentionally broad, the framework further filters them using contextual information in CodeQL. In particular, it checks entity types, local structural relations, and function semantics to remove benign matches that resemble sensitive operations only superficially. For example, it excludes wrapper calls such as \texttt{len(...)}, \texttt{hex(...)}, \texttt{str(...)}, and \texttt{repr(...)}, statistical variables such as \texttt{tokens\_used}, and return expressions from functions without external exposure semantics. This step produces a more precise set of candidate program entities for the subsequent taint analysis.

\subsubsection{Privacy leakage detector based on taint analysis.}

At this stage, \toolname  further determines whether sensitive data can actually flow to potential leakage endpoints, based on the candidate program entities obtained after rule matching and context-aware filtering. The framework first maps candidate program entities to a concrete program expression that can participate in data-flow analysis. For source rules, this mapping binds matched entities (e.g., hardcoded literals, values returned by sensitive file reads, and identifiers with sensitive semantics) to the corresponding expressions that introduce data into execution. For sink rules, it binds matched entities such as logging arguments, HTTP request arguments, and externally visible return expressions to the corresponding output expressions. After this binding step, the framework issues source--sink reachability queries over the CodeQL database.

The analysis validates propagation through three common categories of relations. The first is \emph{assignment propagation}, where a sensitive value is copied or rebound through local assignments. The second is \emph{argument propagation}, where a sensitive value is passed into another function or method call and continues to flow through parameters and return values across call boundaries. The third is \emph{attribute propagation}, where a sensitive value is stored in an object field, later retrieved through attribute access, and then forwarded to subsequent operations. These relations allow the detector to cover both intra-procedural and inter-procedural propagation patterns that commonly appear in MCP servers. A candidate leakage path is reported only when the \toolname can construct a reachable source-sink chain in the program graph.

\section{Evaluation}
\subsection{Dataset and Experiment Setup}
We built the dataset by collecting MCP servers through the GitHub API and several  registries, including \texttt{Smithery Registry}, \texttt{Pulse MCP}, \texttt{Cursor Directory}, \texttt{Awesome MCP}, \texttt{Glama AI}, \texttt{Mcpmarket}, and \texttt{Modelcontextprotocol.io (MCP.io)}. We developed a dedicated crawler to automatically obtain the source code and the associated metadata of each MCP server. In total, we obtained 10,655 MCP servers, covering a wide range of functional categories. To reduce the impact of class imbalance and avoid bias caused by categories with only a small number of samples, we grouped these categories into a single ``Other'' category, making the dataset more balanced and representative. We also assigned a small number of servers to an ``Unknown'' category when their metadata in public registries was missing, incomplete, or inconsistent.

\vspace{5pt}
\noindent\textbf{Execution Environment.} All experiments were conducted on a 64-bit x64-based workstation running Microsoft Windows 11 Pro (Version 10.0.26200, Build 26200). The machine was equipped with an Intel(R) Core(TM) Ultra 7 258V processor at 2.20 GHz, featuring 32 GB of RAM.

\subsection{Performance of \toolname}
\noindent\textbf{Accuracy.} \toolname\ has demonstrated a high detection accuracy rate. To assess the reliability of the detection results, we manually verified the findings produced by \toolname. Specifically, we randomly sampled 200 MCP servers that \toolname\ had identified as exhibiting privacy leaks and subjected them to individual scrutiny. Of these, 192 were confirmed to involve genuine privacy leaks, while only 8 were identified as false positives, yielding a corresponding false positive rate of 4.0\%. Concurrently, we observed no instances of false negatives. These results indicate that \toolname\ is capable of effectively identifying privacy leakage risks across a large-scale population of MCP servers.
\subsection{Empirical Results}

\begin{figure}[t]
  \centering
  \includegraphics[width=1\linewidth]{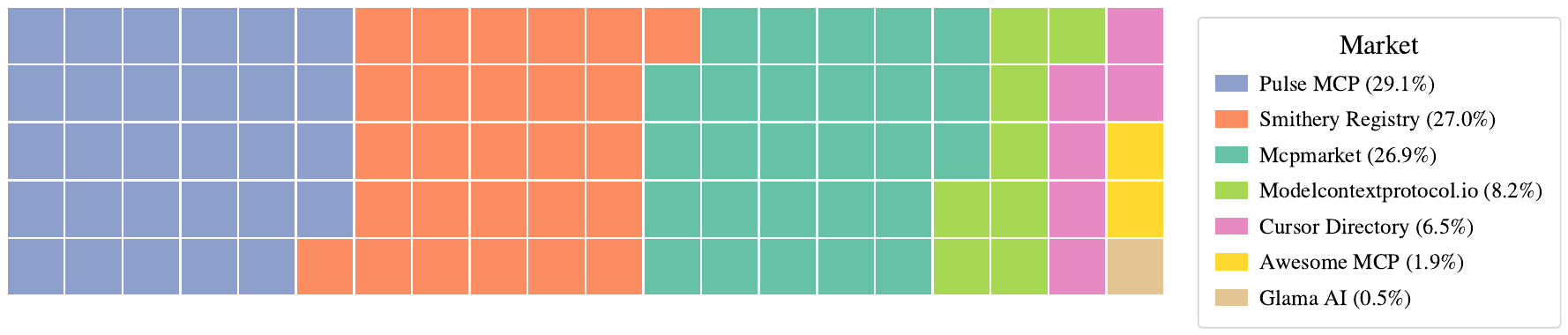}
  \caption{Market contribution of privacy-leaking MCP servers}
  \label{img:MCP_PII_leakage_by_market}
\end{figure}

\noindent\textbf{Market-level distribution of privacy leakage.} Figures~\ref{img:MCP_PII_leakage_rate_by_market} and~\ref{img:MCP_PII_leakage_by_market} present two complementary views of privacy leakage across MCP markets, namely leakage rate and market contribution. From the leakage rate, \texttt{MCP.io} exhibits the highest leakage rate at approximately 18.1\%, followed by \texttt{Smithery Registry} at 15.1\% and \texttt{Pulse MCP} at 13.2\%. \texttt{Mcpmarket} and \texttt{Awesome MCP} are both around 12\%, while \texttt{Cursor Directory} and \texttt{Glama AI} show relatively lower rates, at 9.3\% and 8.8\%, respectively. From the market contribution, however, privacy-leaking MCP servers are mainly concentrated in several major ecosystems. \texttt{Pulse MCP}, \texttt{Smithery Registry}, and \texttt{Mcpmarket} account for 29.1\%, 27.0\%, and 26.9\% of all leakage cases, respectively, whereas \texttt{MCP.io}, \texttt{Cursor Directory}, \texttt{Awesome MCP}, and \texttt{Glama AI} contribute relatively limited shares. These results indicate that some markets exhibit high internal leakage rates even though they contribute only a limited fraction of all leakage cases, whereas large ecosystems dominate the overall leakage distribution because of their scale.

% From the perspective of market contribution as shown in Figure \ref{img:MCP_PII_leakage_by_market}, MCP servers with privacy leakage are primarily concentrated in a small number of major ecosystems. \texttt{Pulse MCP}, \texttt{Smithery Registry}, and \texttt{Mcpmarket} account for 29.1\%, 27.0\%, and 26.9\% of all leakage cases, respectively, indicating that privacy leakage risk is highly concentrated in these three markets. In contrast, \texttt{MCP.io}, \texttt{Cursor Directory}, \texttt{Awesome MCP}, and \texttt{Glama AI} contribute relatively limited shares, revealing a clear head-concentrated distribution overall.

\begin{figure}[t]
  \centering
  \begin{subfigure}[t]{0.45\linewidth}
    \centering
    \includegraphics[width=\linewidth]{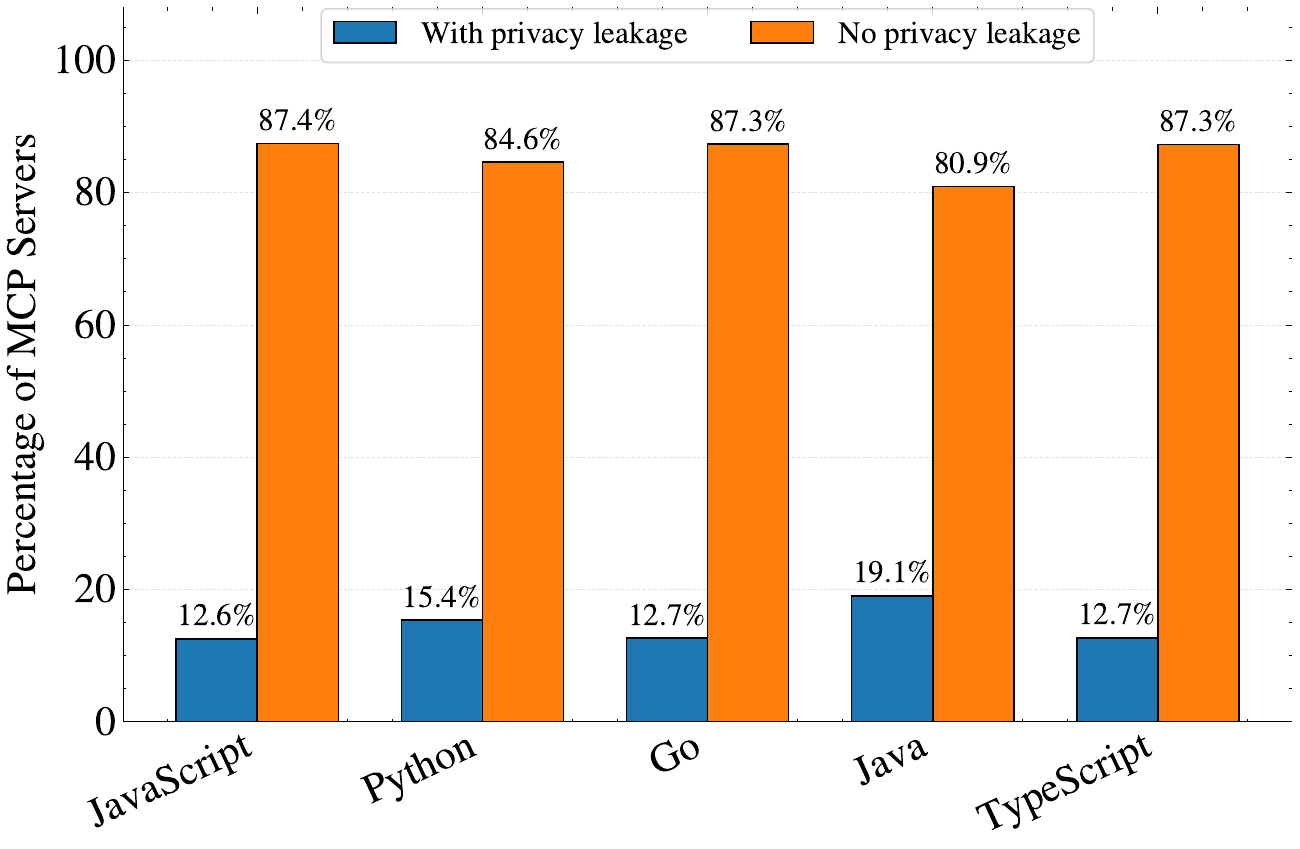}
    \caption{MCP server privacy leakage in various programming languages}
    \label{img:MCP_privacy_by_language_pct}
  \end{subfigure}
  \hfill
  \begin{subfigure}[t]{0.45\linewidth}
    \centering
    \includegraphics[width=\linewidth]{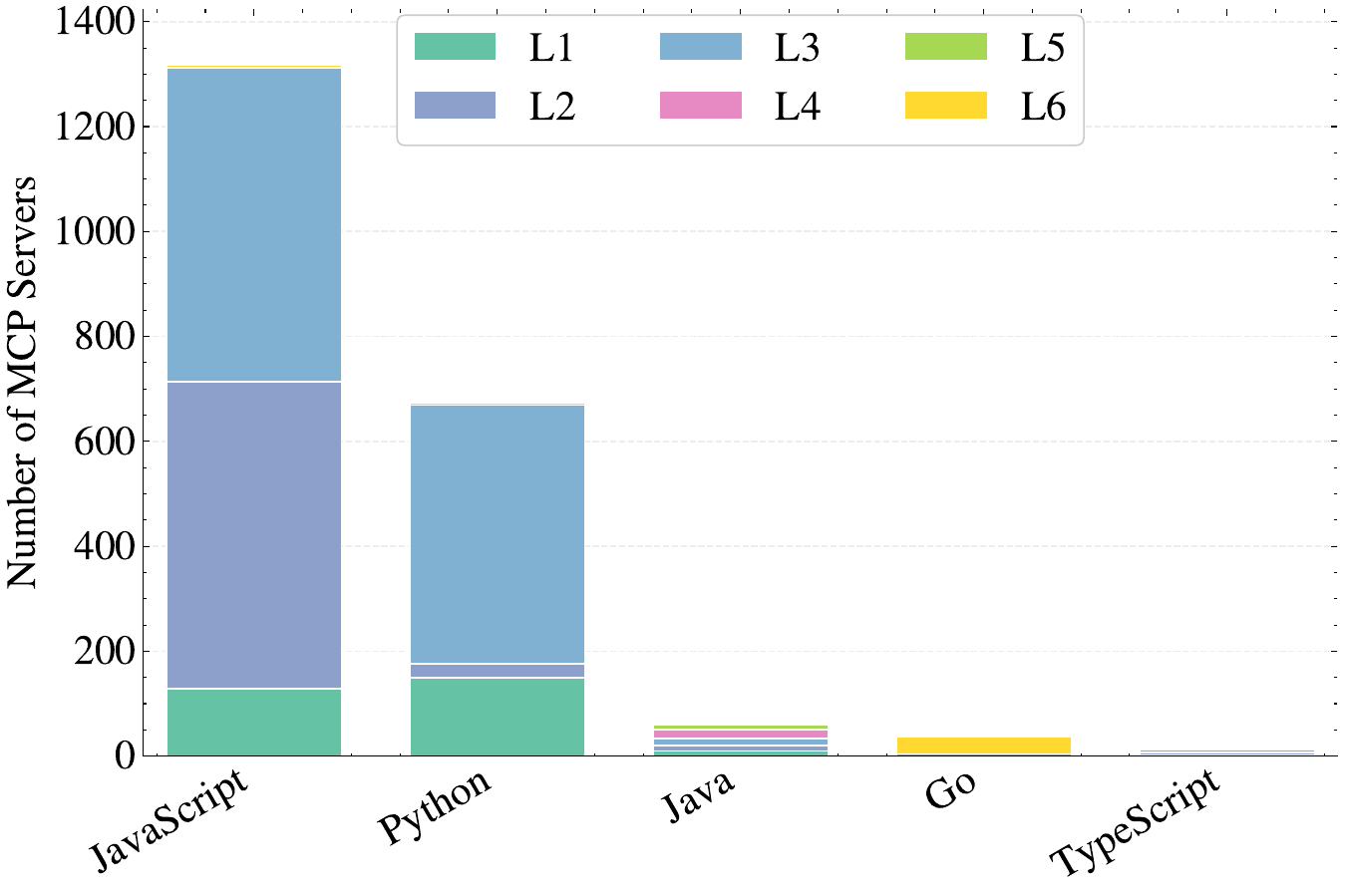}
    \caption{Rule composition of privacy leakage across languages}
    \label{img:MCP_PII_rule_language}
  \end{subfigure}
  \caption{Language-level distribution of privacy leakage and rule composition}
  \label{img:MCP_PII_language_combined}
\end{figure}

\vspace{5pt}
\noindent\textbf{Language-level distribution of privacy leakage and rule composition.} As shown in Figure \ref{img:MCP_privacy_by_language_pct}, it shows the MCP server privacy leakage in various programming language. Under the current rule set, the majority of MCP servers in each language do not exhibit privacy leakage risks; however, the leakage rates vary substantially across languages. Specifically, Java shows the highest leakage rate at 19.1\%, followed by Python at 15.4\%, whereas JavaScript, Go, and TypeScript display relatively similar proportions. These results suggest that privacy leakage risk is not evenly distributed across different language implementations, but instead tends to be more concentrated in particular language ecosystems. Moreover, the leakage rate exceeds 10\% for every language, indicating that this issue is not driven by a small number of anomalous cases, but has already exhibited prevalence across the MCP ecosystem.

\begin{wrapfigure}{r}{0.45\linewidth}
  \centering
  \includegraphics[width=\linewidth]{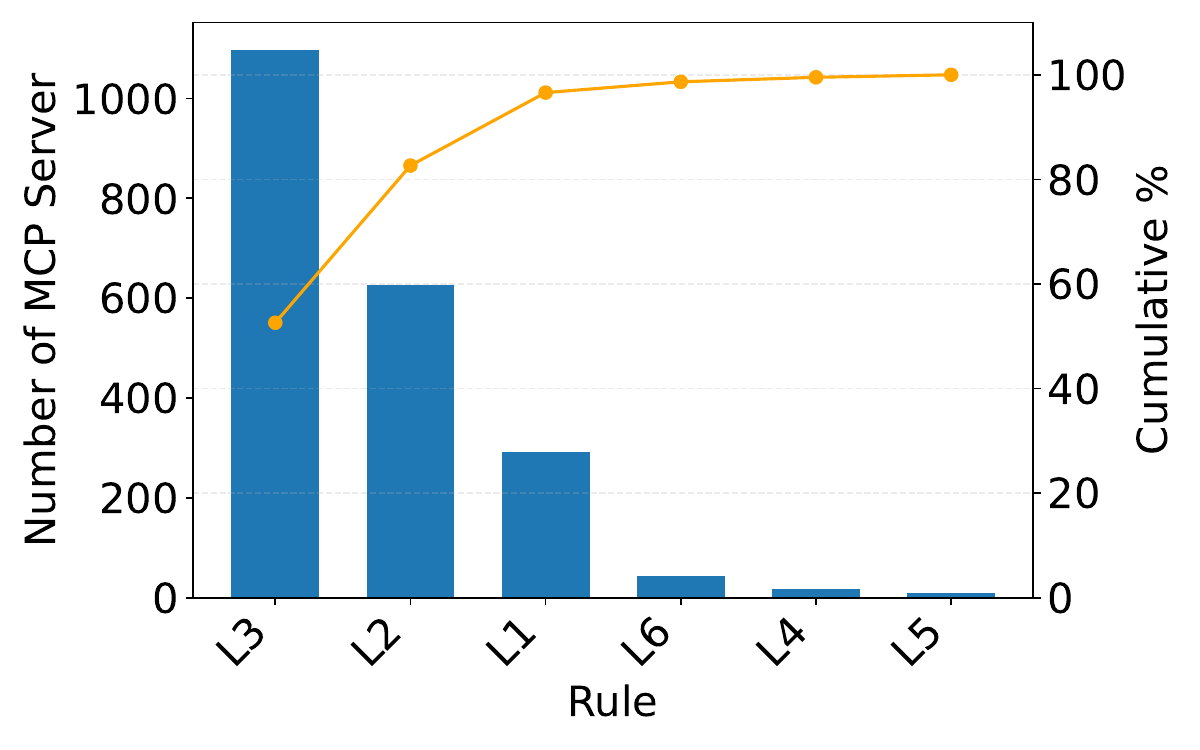}
  \caption{MCP servers with privacy leakage by rules\vspace{-25pt}}
  \label{img:MCP_PII_rule_by_project}
\end{wrapfigure}

As shown in Figure \ref{img:MCP_PII_rule_language}, it shows rule composition of privacy leakage across languages. JavaScript and Python account for the majority of privacy leakage cases, whereas Java, Go, and TypeScript contribute substantially fewer cases, indicating a relatively concentrated language-level distribution. From the perspective of rule composition, L3 and L2 dominate in  JavaScript and Python, suggesting that privacy leakage in the major language is primarily driven by a small number of high-frequency rules. By contrast, L4, L5, and L6 are triggered far less frequently and mainly serve as marginal supplements. These results indicate that privacy leakage risk in the MCP ecosystem exhibits strong concentration in both the language and rule dimensions.

% \begin{figure}[t]
%   \centering
%   \begin{minipage}[t]{0.4\linewidth}
%     \centering
%     \includegraphics[width=\linewidth]{figures/MCP_PII_rule_by_project.pdf}
%     \captionof{figure}{MCP Servers with Privacy Leakage by Rule}
%     \label{img:MCP_PII_rule_by_project}
%   \end{minipage}
%   \hfill
%   \begin{minipage}[t]{0.55\linewidth}
%     \centering
%     \includegraphics[width=\linewidth]{figures/MCP_PII_rule_category.pdf}
%     \captionof{figure}{MCP Servers with Privacy Leakage by Category}
%     \label{img:MCP_PII_rule_category}
%   \end{minipage}
% \end{figure}

\begin{figure*}[t]
        \centering
        \includegraphics[width=\linewidth]{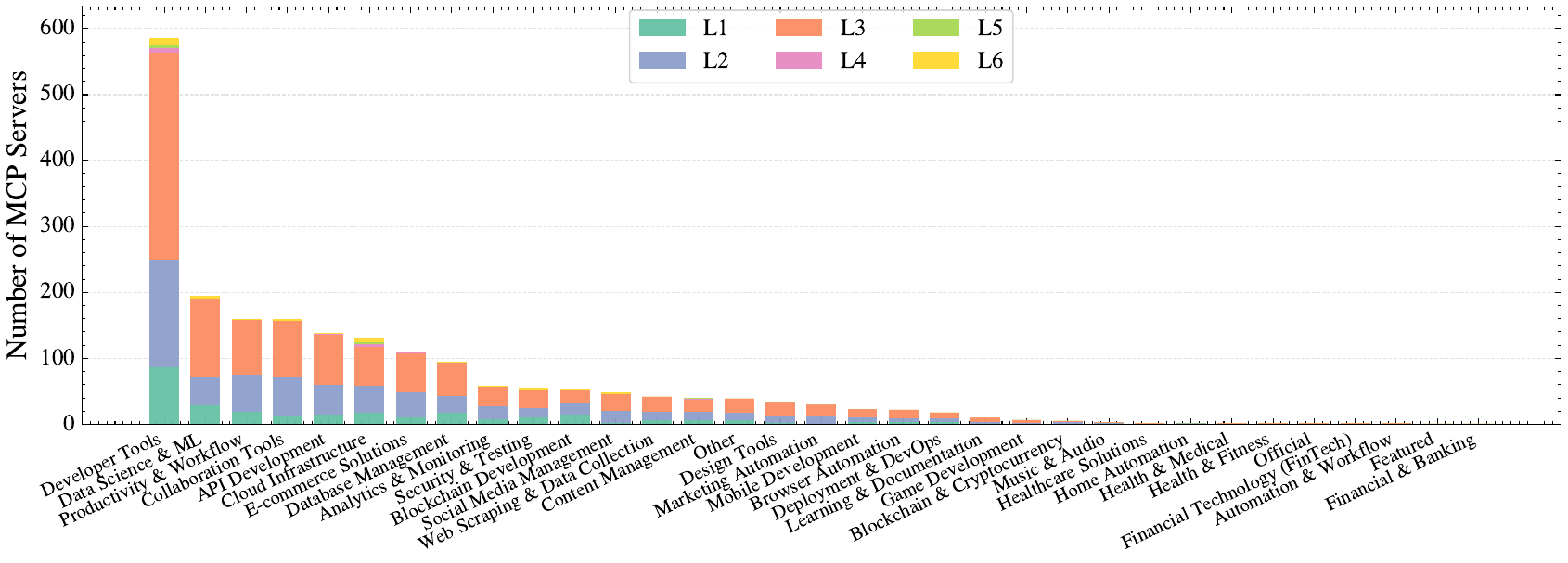}
        \caption{MCP servers with privacy leakage by category}
        \label{img:MCP_PII_rule_category}
\end{figure*}

\vspace{5pt}
\noindent\textbf{Rule-level analysis of privacy leakage.} As shown in Figure \ref{img:MCP_PII_rule_by_project}, the coverage of MCP servers with privacy leakage across different rules exhibits a clear concentration pattern. L3 is the most frequently triggered rule, detecting more MCP servers than any other rule, and when L2 and L1 are further incorporated, the cumulative coverage approaches nearly all privacy leakage cases. In contrast, L6, L4, and L5 are triggered far less frequently and contribute only marginally to the overall coverage, exhibiting a typical long-tail pattern. These results indicate that privacy leakage in the current MCP ecosystem is not driven uniformly by all rules, but is instead concentrated in a small number of high-frequency leakage patterns. Table \ref{tab:table_rule} summarizes the distribution of MCP Servers for each rule within its high-frequency application category. This table indicates that different rules exhibit significant disparities not only in the number of detections, but also in distributed characteristics such as market coverage, language coverage, and primary application categories. Specifically, rules L3, L2, and L1 demonstrate broader distributions, appearing across a wider range of markets and languages. In contrast, rules L4, L5, and L6 show relatively limited coverage, suggesting that the leakage patterns associated with these rules are more concentrated in specific application scenarios.

\begin{table*}[!htb]
\centering
\caption{Coverage of privacy leakage rules}
\label{tab:rule_summary}
{
\begingroup
\setlength{\tabcolsep}{2.0pt}
\renewcommand{\arraystretch}{0.92}
\fontsize{8}{7.2}\selectfont
\resizebox{0.98\textwidth}{!}{%
\begin{tabular}{@{}c c c c p{8.2cm}@{}}
\toprule
\textbf{Rule} & \textbf{Servers} & \textbf{Markets} & \textbf{Languages} & \textbf{Top Categories} \\
\midrule
L1 & 293  & 7 & 5 & Developer Tools, Data Science \& ML \\
L2 & 628  & 7 & 4 & Developer Tools, Collaboration Tools, Productivity \& Workflow \\
L3 & 1110 & 7 & 4 & Developer Tools, Data Science \& ML, Collaboration Tools \\
L4 & 18   & 5 & 2 & Developer Tools, Cloud Infrastructure, E-commerce Solutions \\
L5 & 10   & 4 & 2 & Developer Tools, Cloud Infrastructure, Content Management \\
L6 & 43   & 5 & 3 & Developer Tools, Cloud Infrastructure \\
\bottomrule
\end{tabular}%
}
\endgroup
}
\label{tab:table_rule}
\end{table*}

\vspace{5pt}
\noindent\textbf{Category-level distribution of privacy leakage.} Privacy leakage is distributed across application categories in Figure \ref{img:MCP_PII_rule_category}. Developer Tools account for the largest number of leakage cases, substantially exceeding all other categories. This finding indicates that MCP servers in Developer Tools are more likely to expose privacy risks when handling code artifacts, configuration files, credentials, and external tool interactions. Beyond Developer Tools, categories such as Data Science \& ML, Productivity \& Workflow, Collaboration Tools, API Development, and Cloud Infrastructure also exhibit relatively high numbers of leakage cases, indicating that privacy risks are  prominent in scenarios involving intensive data processing, automated orchestration, and frequent interactions with external resources. 

\begin{table*}[!htb]
\centering
\caption{Distribution of MCP servers by privacy leakage status}
{
\begingroup
\setlength{\tabcolsep}{2.0pt}
\renewcommand{\arraystretch}{0.92}
\fontsize{8}{7.2}\selectfont
\resizebox{0.98\textwidth}{!}{%
\begin{tabular}{@{}p{3.2cm}*{7}{cc}@{}}
\toprule
\multirow{2}{*}{\textbf{Category}} &
\multicolumn{2}{c}{\textbf{Awesome MCP}} &
\multicolumn{2}{c}{\textbf{Cursor Directory}} &
\multicolumn{2}{c}{\textbf{Glama AI}} &
\multicolumn{2}{c}{\textbf{Mcpmarket}} &
\multicolumn{2}{c}{\textbf{MCP.io}} &
\multicolumn{2}{c}{\textbf{Pulse MCP}} &
\multicolumn{2}{c}{\textbf{Smithery Registry}} \\
\cmidrule(lr){2-3}\cmidrule(lr){4-5}\cmidrule(lr){6-7}
\cmidrule(lr){8-9}\cmidrule(lr){10-11}\cmidrule(lr){12-13}\cmidrule(lr){14-15}
& \textbf{L} & \textbf{NL}
& \textbf{L} & \textbf{NL}
& \textbf{L} & \textbf{NL}
& \textbf{L} & \textbf{NL}
& \textbf{L} & \textbf{NL}
& \textbf{L} & \textbf{NL}
& \textbf{L} & \textbf{NL} \\
\midrule
Developer Tools                  
& 3 & 32 & 12 & 172 & - & 6 & 260 & 1893 & 7 & 43 & 57 & 567 & 38 & 311 \\

Data Science \& ML              
& 3 & 25 & 9 & 143 & - & 15 & 21 & 133 & 9 & 52 & 39 & 361 & 52 & 357 \\

Database Management             
& 4 & 22 & 8 & 32 & - & 7 & 7 & 48 & 8 & 27 & 15 & 169 & 24 & 154 \\

API Development                 
& 1 & 12 & 4 & 30 & - & 1 & 18 & 99 & 5 & 22 & 36 & 169 & 23 & 128 \\

Web Scraping \& Data Collection 
& 1 & 17 & 5 & 58 & - & 3 & 9 & 135 & 1 & 13 & 7 & 100 & 10 & 132 \\

Cloud Infrastructure            
& - & 11 & 6 & 37 & 3 & 12 & 9 & 68 & 11 & 30 & 27 & 122 & 24 & 90 \\

Collaboration Tools             
& - & 11 & 8 & 53 & - & 7 & 1 & 2 & 7 & 23 & 39 & 116 & 41 & 139 \\

Productivity \& Workflow        
& 2 & 8 & 10 & 59 & - & 2 & - & 3 & 7 & 18 & 17 & 90 & 49 & 157 \\

E-commerce Solutions            
& 3 & 5 & 8 & 37 & 1 & 5 & - & - & 6 & 15 & 24 & 83 & 31 & 118 \\

Analytics \& Monitoring         
& - & 8 & 2 & 28 & - & - & 22 & 138 & - & 17 & 6 & 42 & 7 & 50 \\

Security \& Testing             
& - & 9 & 2 & 28 & 2 & 2 & 3 & 5 & 1 & 14 & 14 & 91 & 10 & 64 \\

Content Management              
& - & 4 & 3 & 29 & - & 2 & 1 & 6 & 3 & 13 & 6 & 84 & 9 & 61 \\

Social Media Management         
& 1 & 2 & 5 & 20 & - & 1 & 4 & 9 & 1 & 8 & 12 & 41 & 9 & 55 \\

Browser Automation              
& 2 & 6 & 1 & 12 & - & 5 & - & 1 & 1 & 1 & 6 & 41 & 5 & 49 \\

Design Tools                    
& - & 1 & 1 & 15 & - & - & 1 & 2 & 2 & 5 & 9 & 25 & 9 & 55 \\

Mobile Development              
& - & 3 & 1 & 10 & - & - & - & 1 & 1 & 7 & 4 & 30 & 10 & 48 \\

Blockchain Development          
& 1 & 1 & 2 & 9 & - & - & - & - & 6 & 2 & 14 & 45 & 8 & 25 \\

Game Development                
& - & 3 & - & 11 & - & 1 & - & - & - & 4 & 3 & 49 & 2 & 26 \\

Unknown                         
& 4 & 1 & - & 2 & - & - & - & 3 & 39 & 196 & 56 & 362 & 2 & 24 \\

Other                           
& - & - & - & 1 & - & 1 & 23 & 157 & - & - & - & 1 & - & 10 \\
\bottomrule
\end{tabular}%
}
\endgroup
}
\label{table_category}
\end{table*}

Table \ref{table_category} presents the counts of MCP Servers exhibiting privacy leaks (L) versus those with no detected leaks (NL) across the top 20 categories; thus, it reflects not only the distribution of leaks but also the overall scale of the ecosystem within each category and market. Categories such as Developer Tools, Data Science \& ML, API Development, and Database Management consistently exhibit high counts of both L and NL servers across multiple markets. This suggests that privacy leaks are not randomly distributed across various application scenarios, but rather occur more frequently in those categories and markets that possess large-scale MCP servers.

\subsection{Case Studies}

\begin{figure*}[t]
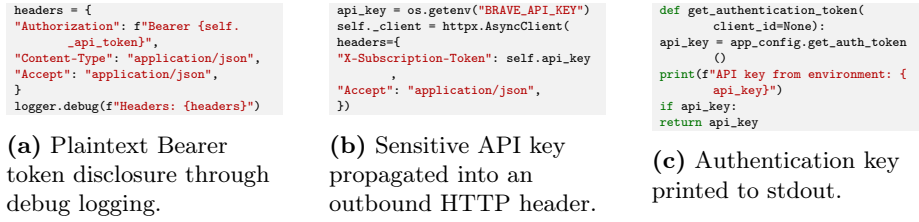

\centering

\begin{minipage}[t]{0.30\textwidth}
\centering
\begin{lstlisting}[style=mcppython]
headers = {
"Authorization": f"Bearer {self._api_token}",
"Content-Type": "application/json",
"Accept": "application/json",
}
logger.debug(f"Headers: {headers}")
\end{lstlisting}
\vspace{3pt}
{\raggedright\small \textbf{(a)} Plaintext Bearer token disclosure through debug logging.\par}
\end{minipage}
\hfill
\begin{minipage}[t]{0.30\textwidth}
\centering
\begin{lstlisting}[style=mcppython]
api_key = os.getenv("BRAVE_API_KEY")
self._client = httpx.AsyncClient(
headers={
"X-Subscription-Token": self.api_key,
"Accept": "application/json",
})
\end{lstlisting}
\vspace{3pt}
{\raggedright\small \textbf{(b)} Sensitive API key propagated into an outbound HTTP header.\par}
\end{minipage}
\hfill
\begin{minipage}[t]{0.30\textwidth}
\centering
\begin{lstlisting}[style=mcppython]
def get_authentication_token(client_id=None):
api_key = app_config.get_auth_token()
print(f"API key from environment: {api_key}")
if api_key:
return api_key
\end{lstlisting}
\vspace{3pt}
{\raggedright\small \textbf{(c)} Authentication key printed to stdout.\par}
\end{minipage}

\vspace{4pt}
\caption{Representative privacy leakage patterns in real-world MCP servers}
\label{fig:case-studies-privacy-leakage}
\end{figure*}

\noindent\textbf{Case Study 1: Bearer token disclosure via debug logging.}
 Figure~\ref{fig:case-studies-privacy-leakage}(a) shows a representative leakage pattern in a real-world MCP server. The program first constructs an authenticated HTTP header containing a Bearer token and then directly writes the entire \texttt{headers} dictionary to a debug log. Since the \texttt{Authorization} field contains \texttt{self.\_api\_token}, the complete token is exposed at a logging sink. Once debug logging is enabled, the token may appear in console output, local log files, or centralized logging infrastructures, from which an attacker could recover and reuse it for unauthorized API invocation.

\vspace{5pt}
\noindent\textbf{Case Study 2: Sensitive API key propagation to an outbound HTTP header.}
As illustrated in  Figure~\ref{fig:case-studies-privacy-leakage}(b), the MCP server first loads an API key from the environment and subsequently inserts it into an outbound HTTP request header. This is a valid privacy leakage path, because a privacy-sensitive value leaves the local execution boundary and becomes part of an external request. Propagating sensitive credentials to external request parameters or request headers inherently poses a risk of privacy leakage, as these credentials may be exposed and misused within remote services or logging systems.

\vspace{5pt}
\noindent\textbf{Case Study 3: Plaintext printing of an authentication key.}
Figure~\ref{fig:case-studies-privacy-leakage}(c) presents another concrete leakage pattern, where an authentication key retrieved from configuration or environment context is printed directly to standard output. Although the value is not immediately exfiltrated through a network request, it becomes observable through terminal output, system logs. This case demonstrates that privacy leakage in MCP servers can arise not only from explicit external transmission, but also from local outputs during execution.

\section{Related Work}

\noindent{\bf MCP Security.} MCP is a standardized interface that connects LLMs with external tools by structuring inputs and coordinating multi-source information \cite{karimova2025model,11069802,11105280,li2026don,hou2025model}. However, recent studies have shown that if an insecure MCP is connected, it can become an exploit path for attackers to control LLM behavior, inject malicious instructions, steal assets, and achieve remote code execution (RCE) \cite{liu2024demystifying,dong2025philosopher,guo2025systematic,kumar2025mcp}. Specifically, untrusted MCP data sources or tool responses can lead to serious risks such as prompt injection \cite{liu2025datasentinel,liu2024formalizing,shen2024anything}, cross-service prompt stealing \cite{yangprsa}, demystifying RCE manipulation of executors \cite{liu2024demystifying,radosevich2025mcp}, and trojanizing plugins \cite{dong2025philosopher}. Furthermore, the complexity of responsibility attribution introduced by long contexts requires enhanced traceability in MCP design \cite{wangtracllm}. 

\noindent{\bf MCP Privacy.} Recent studies indicate that MCP increases the risk of privacy leakage by enabling the flow of sensitive information across various tools, services, and execution contexts \cite{croce2025trivial,zhao2025mind,fang2025we,sun2025msa}. Existing work suggests that such risks may stem from cross-tool interactions and parasitic toolchains \cite{croce2025trivial,zhao2025mind}, or from the integration of untrusted third-party MCP services \cite{fang2025we}. Concurrently, some MCP-based systems have begun to incorporate mechanisms,such as access control, encryption, and audit logs, to mitigate the risk of privacy exposure in highly sensitive application scenarios \cite{shehab2025agentic,ehtesham2025enhancing}. These research findings demonstrate that privacy security within an MCP environment is closely intertwined with the protocol's orchestration model and its cross-context data access patterns. our work focuses on privacy leakage detection in MCP servers and proposes a systematic static analysis approach tailored to the MCP ecosystem.

\section{Conclusion}

In this paper, we investigated privacy leakage risks in MCP servers through \toolname, a cross-language static analysis framework. By combining unified program representation, context-aware rule matching, and taint analysis, \toolname identifies privacy leakage in real-world MCP deployments. Evaluation on 10,655 MCP servers shows that privacy leakage is widespread across the MCP ecosystem, with non-trivial leakage rates across all languages and clear concentration in several major markets, dominant rules, and application categories. Case studies further confirm concrete leakage in practice, including leaked Bearer tokens, propagated API keys, and plaintext authentication credentials. Overall, our findings show that privacy leakage has become a significant security concern in MCP servers and highlight the need for systematic detection across the MCP ecosystem.

% \begin{credits}
% \subsubsection{\ackname} A bold run-in heading in small font size at the end of the paper is
% used for general acknowledgments, for example: This study was funded

% \subsubsection{\discintname}
% It is now necessary to declare any competing interests or to specifically
% state that the authors have no competing interests. Please place the
% statement with a bold run-in heading in small font size beneath the
% (optional) acknowledgments\footnote{If EquinOCS, our proceedings submission
% system, is used, then the disclaimer can be provided directly in the system.},
% for example: The authors have no competing interests to declare that are
% relevant to the content of this article. Or: Author A has received research
% grants from Company W. Author B has received a speaker honorarium from
% Company X and owns stock in Company Y. Author C is a member of committee Z.
% \end{credits}
\section*{Open Science}
We provide all source code necessary to evaluate the core contributions of this paper through a repository hosted at https://anonymous.4open.science/r/mcp\_priv/.
\bibliographystyle{splncs04}
\bibliography{bib/ref}
\end{document}